\documentclass[twocolumn,showpacs,superscriptaddress]{revtex4-1}
\usepackage{hyperref}
\hypersetup{
  colorlinks=true,        
  linkcolor=blue,         
  citecolor=cyan,         
}
\usepackage[T1]{fontenc}

\usepackage{graphicx}
\usepackage{dcolumn}
\usepackage{bm}
\usepackage{color}
\usepackage{enumitem}
\usepackage{amsmath}
\usepackage{amssymb}
\usepackage{orcidlink}
\usepackage{graphicx} 
\usepackage{subcaption}
\usepackage{caption}
\usepackage{booktabs, tabularx}
\usepackage{dcolumn}
\usepackage{makecell}
\usepackage{multirow}
\usepackage{soul}
\setstcolor{red}
\usepackage{xcolor}

\begin{document}
\title{Optical Signatures of a Black Hole in Effective Quantum Gravity}

\author{Jonibek Khasanov}
\email{khasanovjonibek775@gmail.com}
\affiliation{Institute of Fundamental and Applied Research, National Research University TIIAME, Kori Niyoziy 39, Tashkent 100000, Uzbekistan}
\affiliation{University of Tashkent for Applied Sciences, Str. Gavhar 1, Tashkent 100149, Uzbekistan}
\affiliation{Information Technologies Department, 
National Research University TIIAME, Kori Niyoziy 39, Tashkent 100000, Uzbekistan}

\author{Xuwei Zhang} 
\email{24B911021@stu.hit.edu.cn}
\affiliation{School of Physics, Harbin Institute of Technology, Harbin 150001, People’s Republic of China}

\author{Mirzabek Alloqulov}
\email{malloqulov@gmail.com}
\affiliation{School of Physics, Harbin Institute of Technology, Harbin 150001, People’s Republic of China}
\affiliation{Department of Physics, New Uzbekistan University, Movarounnahr str. 1, Tashkent 100000, Uzbekistan}
\affiliation{Tashkent State Technical University, Tashkent 100095, Uzbekistan}

\author{Sanjar Shaymatov}
\email{sanjar@astrin.uz}
\affiliation{Institute of Fundamental and Applied Research, National Research University TIIAME, Kori Niyoziy 39, Tashkent 100000, Uzbekistan}
\affiliation{National University of Uzbekistan, Tashkent 100174, Uzbekistan}

\author{Chen Zhou}
\email{chenzhou@hit.edu.cn}
\affiliation{School of Physics, Harbin Institute of Technology, Harbin 150001, People’s Republic of China}

\author{Chengxun Yuan}
\email{yuancx@hit.edu.cn}
\affiliation{School of Physics, Harbin Institute of Technology, Harbin 150001, People’s Republic of China}

\date{\today}

\begin{abstract}

In this article, we present an investigation of observational signatures of a black hole (BH) in effective quantum gravity (EQG), particularly gravitational weak lensing and shadow analyses with the assumption that the BH is surrounded by uniform and non-uniform plasma. First, we review the spacetime by comparing it with well-known BH spacetimes such as Schwarzschild and Reissner--Nordstr\"{o}m BHs. Subsequently, we explore the effect of the uniform and non-uniform plasma on the BH shadow. The results indicate that shadow radii decrease under the influence of the BH charge and both uniform and non-uniform plasma parameters. In addition, we obtain theoretical constraints on the spacetime parameters using observational data from the Event Horizon Telescope (EHT) collaboration. Finally, we study gravitational weak lensing around the BH in EQG surrounded by uniform and non-uniform plasma. It was found that increasing the uniform plasma frequency leads to a rise in the deflection angle and vice versa for the non-uniform plasma.  
\end{abstract}

\maketitle

\section{Introduction} 

General relativity (GR) provides the standard theoretical framework for describing the gravitational interaction and has successfully passed a wide range of experimental and observational tests. Black holes (BHs) are among its most remarkable predictions, providing natural laboratories for investigating gravity in the strong-field regime. Nevertheless, the classical description of BHs encounters a fundamental difficulty at the spacetime singularity, where curvature becomes divergent, and the classical theory is expected to cease being applicable. This problem has motivated the development of quantum theories of gravity and, in particular, effective approaches in which quantum-gravitational effects are incorporated into BH geometry. Loop quantum gravity (LQG) is one of the approaches that has inspired several effective BH models in which the classical singularity is replaced by a regular high-curvature region or a transition surface~\cite{Ashtekar_2005, doi:10.1142/S0217751X08039980, Rovelli_2015, Modesto_2010, Ashtekar_2018, Ashtekar_2020, Ashtekar2018, Corichi_2016}. Such constructions can retain an asymptotically flat exterior while modifying the geometry in the strong-field region, thereby providing a framework in which possible quantum-gravitational signatures can be investigated observationally.

Among the various effective quantum-gravity constructions, polymerized BH models provide a useful phenomenological setting for examining deviations from classical BH spacetimes~\cite{Bodendorfer_2021, Bodendorfer2021, Elizaga_Navascu_s_2022, Elizaga_Navascu_s2022, Alonso_Bardaji_2022, Moreira_2023}. In particular, effective charged BH solutions obtained through polymerization of the Reissner-Nordstr\"om Hamiltonian exhibit quantum corrections to the radial sector while preserving the familiar asymptotic behavior and horizon structure in an appropriate parameter regime. The corresponding transition surface replaces the classical singular region, and the quantum-corrected solution approaches the Reissner-Nordstr\"om spacetime in the classical limit~\cite{Borges_2024}. Recent studies have investigated this spacetime through different phenomena~\cite{dePaula:2026vcx, Mohammadi:2026xvg, Kumar:2025ueq, Sekhmani:2025bsi, Menezes:2025mfa}. These features make such geometries especially suitable for studying observables associated with photon propagation and strong gravitational fields. Moreover, because the quantum correction can modify the trajectories of null particles, optical phenomena provide a promising avenue for testing departures from classical BH predictions.

The recent development of BH imaging has provided a particularly important observational motivation for studying such effects. The Event Horizon Telescope (EHT) observations of the compact objects at the centers of M87* and the Milky Way have demonstrated that the characteristic emission structure surrounding a BH can be used to probe the strong-field gravitational environment~\cite{Akiyama2019b, Akiyama2019c, Akiyama2022}. The BH shadow, together with its associated photon ring and lensing features, is determined by the propagation of photons near the unstable circular photon orbit and is therefore sensitive to the underlying spacetime geometry. Consequently, modifications of the metric induced by quantum-gravitational effects may leave observable imprints in the apparent size and structure of the shadow. Comparing these theoretical predictions with EHT observations can therefore provide constraints on parameters describing deviations from classical BH geometries. The BH shadow has been studied extensively in various spacetimes~\cite{Raza:2025ohk, Ali:2024ssf, Kumar:2023jgh, Perlick:2017fio, Perlick:2023znh, Feleppa:2024vdk}. 

However, the propagation of photons around an astrophysical BH is generally not expected to occur in vacuum. Matter surrounding the compact object can form a tenuous plasma, and the interaction between electromagnetic waves and the plasma changes the effective propagation of light. In a dispersive plasma, the photon trajectory depends not only on the spacetime geometry but also on the local plasma frequency together with the photon frequency. This introduces an additional modification to the BH shadow as well as gravitational lensing observables. For a spatially homogeneous plasma, the plasma contribution is characterized by a constant plasma frequency, whereas a radially varying plasma density produces a non-uniform plasma whose influence changes along the photon trajectory. These two cases provide useful limiting descriptions for examining how environmental effects can compete with or enhance modifications produced by the underlying gravitational geometry. There is a substantial body of literature on gravitational lensing as well as the BH shadow in the presence of plasma~\cite{Bisnovatyi-Kogan2010a, Tsupko12, Cunha20a, 2021PDU....3200798B, Javed22GRG, Jafarzade21a, Atamurotov22, Ovgun19WG, Rahvar19MNRAS, Izmailov19MNRAS, Al-Badawi24EPJCa, Al-Badawi24EPJCb, Bisnovatyi-Kogan:2017kii, Atamurotov:2021cgh, 2013Ap&SS.346..513M, Bisnovatyi-Kogan:2010flt}.

Motivated by these considerations, in this paper we investigate the optical properties of the BH in effective quantum gravity (EQG), focusing on its shadow and gravitational weak lensing characteristics in the presence of both uniform and non-uniform plasma. We first review the BH geometry and discuss its relation to familiar classical solutions, particularly the Schwarzschild and Reissner-Nordstr\"om spacetimes. We then derive the conditions governing photon propagation and examine how the BH parameters and plasma properties affect the shadow radius. Furthermore, we confront the theoretical shadow predictions with the available EHT observations to obtain constraints on the relevant spacetime parameters. Finally, we investigate the weak gravitational deflection angle in EQG spacetime for both uniform and non-uniform plasma distributions. 

The article is structured in the following way: Section~\ref{sec:metric} demonstrates the spacetime of the BH in EQG together with the photon dynamics. In Section~\ref{sec:shadow}, we investigate the effect of the uniform and non-uniform plasma on the BH shadow, and we constrain BH charge and plasma frequency by using the EHT results. Section~\ref{sec:weak} is devoted to the gravitational weak lensing around the BH in EQG surrounded by uniform and non-uniform plasma, while Section~\ref{sec:mag} represents the magnification of gravitationally lensed images. We summarize our key results together with their implications in Section~\ref{sec:summary}. In this paper, we adopt the metric signature and geometric units as $(-,+,+,+)$ and $G=c=1$, respectively.

\section{SPACETIME METRIC}

\label{sec:metric}

In this work, we consider a static and axially symmetric BH geometry in the framework of effective quantum gravity (EQG), described by the line element \cite{Borges_2024}:
\begin{equation} \label{metric1}
    ds^2=-A(r)\,dt^2+B(r){dr^2}+C(r)(d\theta^2+\sin^2\theta\,d\phi^2),
\end{equation}
with
\begin{equation}
\begin{aligned}
A(r) &= \left(1-\frac{2M}{r}+\frac{Q^2}{r^2}\right),\\
B(r)&=\left[A(r)\left(1-\frac{r_0}{M}g(r)\right)\right]^{-1},\\
C(r)&=r^2,
\end{aligned}
\end{equation}
where $M$, $r_0$ and $b_0$ denote the BH mass, the transition-surface radius, and the parameter related to polymerization, respectively. Here, the function $g(r)$ reads~\cite{Borges_2024}
\begin{equation}
g(r)=\frac{2Mr-Q^2}{r^2\left(1+\sqrt{1-\frac{b_0^2Q^2}
{(b_0^2-1)^2M^2}}
\right)},
\label{eq:gr}
\end{equation}
and the radius of the transition surface $r_0$ is given in the following form
\begin{equation}
r_0=
\frac{(b_0^2-1)M}{b_0^2}
\left(1+\sqrt{1-\frac{b_0^2Q^2}
{(b_0^2-1)^2M^2}}
\right),
\label{eq:r0}
\end{equation}
where the electric charge, $Q$, follows the following constraint:
\begin{equation}
|Q| \leq \frac{\sqrt{b_0^2-1}}{b_0}\,M,
\qquad \text{for} \qquad b_0>1.
\end{equation}
The parameter $b_0$ depends on the polymerization parameter $\delta_b$ and the Barbero-Immirzi parameter $\gamma$ through the relation $b_0 = \sqrt{1+\gamma\delta_b^2}$.
In the classical limit, when $\delta_b \to 0$ and therefore $b_0 \to 1$, the metric (\ref{metric1}) reduces to the standard RN BH solution.

The polimerisation parameter $\delta_b$ can be related to the mass $M$ and charge $Q$ of the BH using the minimum field condition based on LQG. \cite{Modesto_2010}
\begin{equation}
    4\pi r_0^2 = 4\pi\sqrt{3}\,\gamma .
\end{equation}

as
\begin{equation}
    \delta_b^2 =
\frac{\sqrt{3}}
{\gamma\left(2\times 3^{1/4}\sqrt{\gamma M}-Q^2-\sqrt{3}\gamma\right)}.
\end{equation}
In this work, we considered the Barbero-Immirzi parameter to be equal to $\sqrt{3}/6$ \cite{Pigozzo_2021}.

The Eq.~\eqref{metric1} defines the effective black hole spacetime in which the influence of EQG corrections on the underlying geometry can be examined. To characterize the motion of photons in this gravitational background, we adopt the geodesic Lagrangian formalism and obtain the equations of motion through the Euler–Lagrange equations as

\begin{eqnarray}
    \mathcal{L} &=& \frac{1}{2} g_{\sigma\lambda} \dot{x}^\sigma \dot{x}^\lambda= \\&=& \nonumber\frac{1}{2} \left[ - A(r) \dot{t}^2 + B(r) \dot{r}^2 +r^2d\dot{\theta}^2+ r^2\sin^2\theta \dot{\phi}^2 \right]\ ,
\end{eqnarray}  
\begin{equation}\label{euler}
    \frac{d}{d\tau}\Big(\frac{\partial \mathcal{L}}{\partial  \dot{x}^\sigma} \Big)-\frac{\partial \mathcal{L}}{\partial{x}^\sigma}=0\, .
\end{equation}
In this expression, $\dot{x}^\sigma=dx^\sigma/d\tau$ ($\sigma = 0,1,2,3$) represents the components of the vector tangent to the particle trajectory, while $\tau$ is the corresponding affine parameter. Using the Euler-Lagrange equations for the equatorial plane ($\theta=\pi/2$) and null geodesic condition $(g_{\sigma\lambda} \dot{x}^\sigma \dot{x}^\lambda = 0)$, we obtain conserved quantities for photons such as energy $E$ and momentum $L$, as well as the radial equation of motion.
  
\begin{eqnarray}
    p_t\equiv\frac{\partial \mathcal{L}}{\partial \dot{t}}  &=& -A(r) \dot{t}= -E\, ,\\
\label{p}
 p_\phi\equiv\frac{\partial \mathcal{L}}{\partial \dot{\phi}}  &=& r^2  \dot{\phi}=L\, , \\
\label{energyangular}
\dot{r}^2 &=& \frac{E^2-V_{\rm eff}}{A(r) B(r)}\, ,\label{radialeq}
\end{eqnarray}
In this expression, $p_\sigma\equiv\partial \mathcal{L}/\partial \dot{x}^\sigma$ represents generalized momentum. The effective potential $V_{\mathrm{eff}}(r)$ is expressed as follows:

\begin{equation}
  V_{\rm eff}= A(r)\frac{L^2}{r^2}= \left(1-\frac{2m}{r}+\frac{Q^2}{r^2}\right)\frac{L^2}{r^2}\, .
\end{equation}
Using the condition $\left. \frac{dV_{\rm eff}}{dr} \right|_{r=r_{\rm ph}}=0$, we find the numerical value of the photon-sphere radius $r_{\rm ph}$. By imposing the conditions for an unstable circular photon orbit, $\dot{r}=0$ and $r=r_{\rm ph}$, on the radial equation of motion in Eq.~\eqref{radialeq}, we determine the critical impact parameter $b_c=L/E$. For a distant observer in a static and axial symmetric spacetime, the apparent shadow radius coincides with the critical impact parameter, $R_{\rm sh}=b_c$.  Based on this condition, we find the shadow radius:

\begin{equation}
R_{\rm sh}=b_c = \frac{r_{\rm ph}}{\sqrt{A(r_{\rm ph})}}.
\end{equation}

\begin{figure}
    \centering
    \includegraphics[scale=0.5]{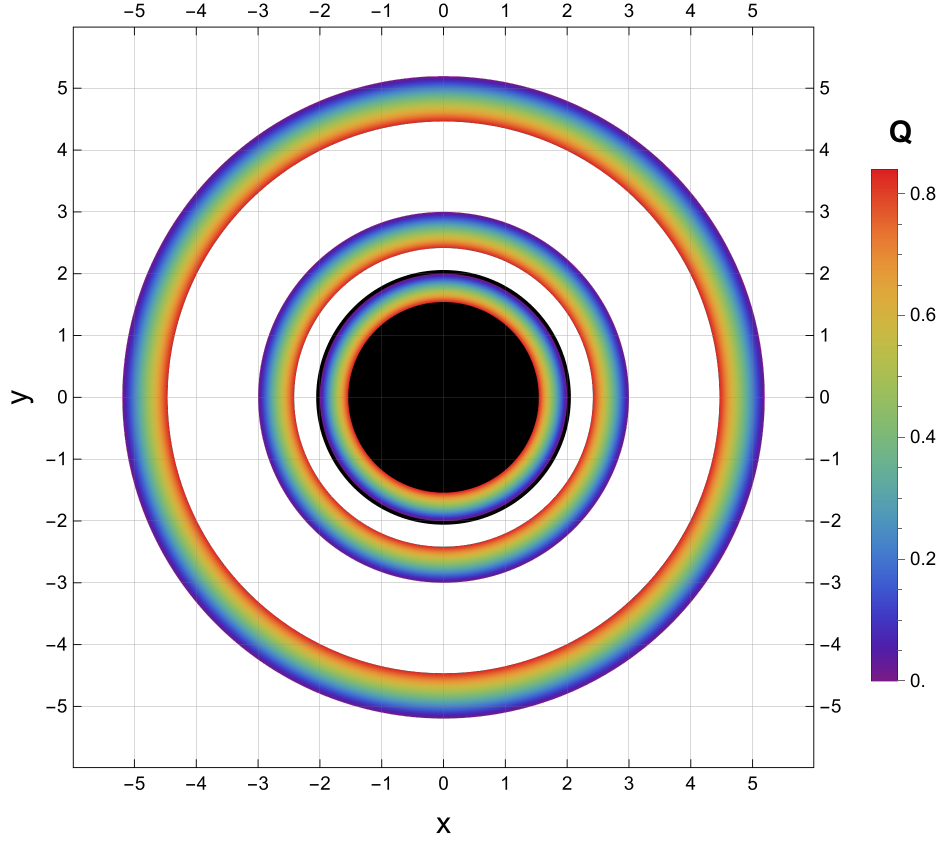}
    \caption{The plot shows the dependence of the event horizon, photon sphere, and shadow radii on the charge of the BH in EFG.} 
    \label{fig1}
\end{figure}

In Fig.~\ref{fig1}, we show the main radii of the BH for different values of the electric charge $Q$. The central black region describes the interior of the event horizon. The color bar on the right shows the values of $Q$. The inner colored ring represents the event horizon radius, while the middle and outer rings show the photon-sphere radius and shadow radius, respectively. As the BH charge increases, all these radii become smaller. This shows that the electric charge changes the spacetime geometry around the BH and reduces the size of the photon region and shadow.

\section{BLACK HOLE SHADOW IN THE PLASMA ENVIRONMENT}
\label{sec:shadow}

We can describe the motion and behavior of photons in a plasma medium using this Hamiltonian~\cite{Synge1960}
\begin{equation}
\mathcal{H}(x^{\alpha}, p_{\alpha}) = \frac{1}{2} 
\left[ g^{\alpha\beta} p_{\alpha} p_{\beta} 
- \left(n^{2}-1\right) \left(p_{\beta} u^{\beta}\right)^{2} \right]\ . \label{eqH}
\end{equation}
Here, $x^\alpha$ shows the spacetime coordinates, $u^\beta$ is the four-velocity of the plasma, and $p_\alpha$ is the four-momentum of the photon. The quantity $n$ is the refractive index of the plasma and is defined as $n=\omega/k$, where $k$ is the wave number. It can be written as follows:

\begin{equation}
    n^2=1-\frac{w^2_{p}}{w^2}\ ,\label{eqn}
\end{equation}
where $\omega_p$ and $\omega$ represent the plasma and photon frequencies, respectively. We define these frequencies using the following expressions:

\begin{equation}
\begin{aligned}
\omega_p^2(x^\alpha) &= \frac{4\pi e^2 N(x^\alpha)}{m_e}, \\ 
\omega(r) &= \frac{\omega_0}{\sqrt{A(r)}},
\qquad \omega_0 = \mathrm{const}.
\end{aligned}
\end{equation}
The quantities $m_e$, $e$, and $N$ correspond to the electron mass, electron charge, and electron number density, respectively.

\begin{figure*}[!htb]
 \centering
\includegraphics[scale=0.55]{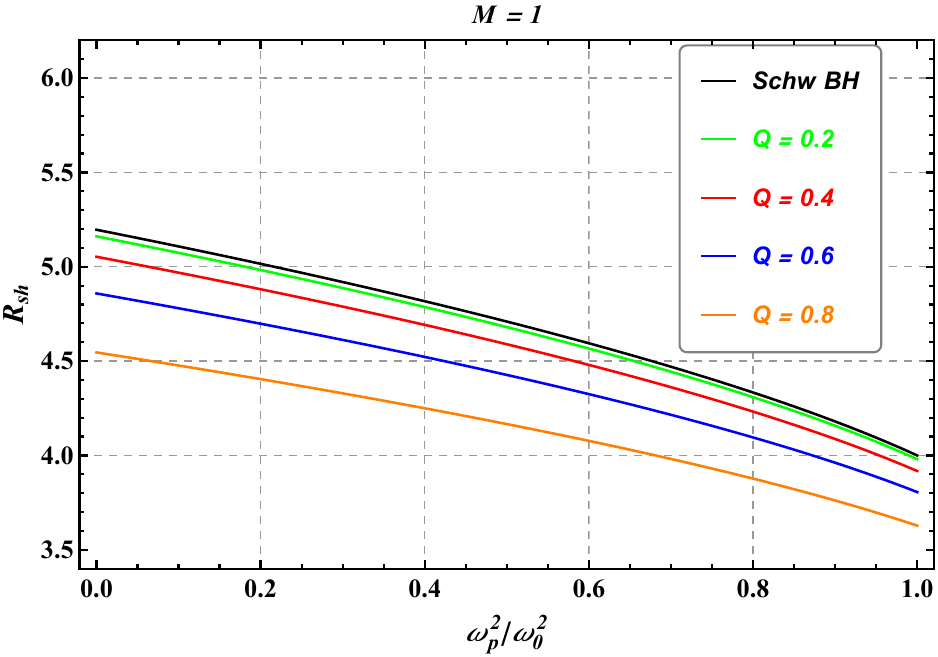}
\includegraphics[scale=0.55]{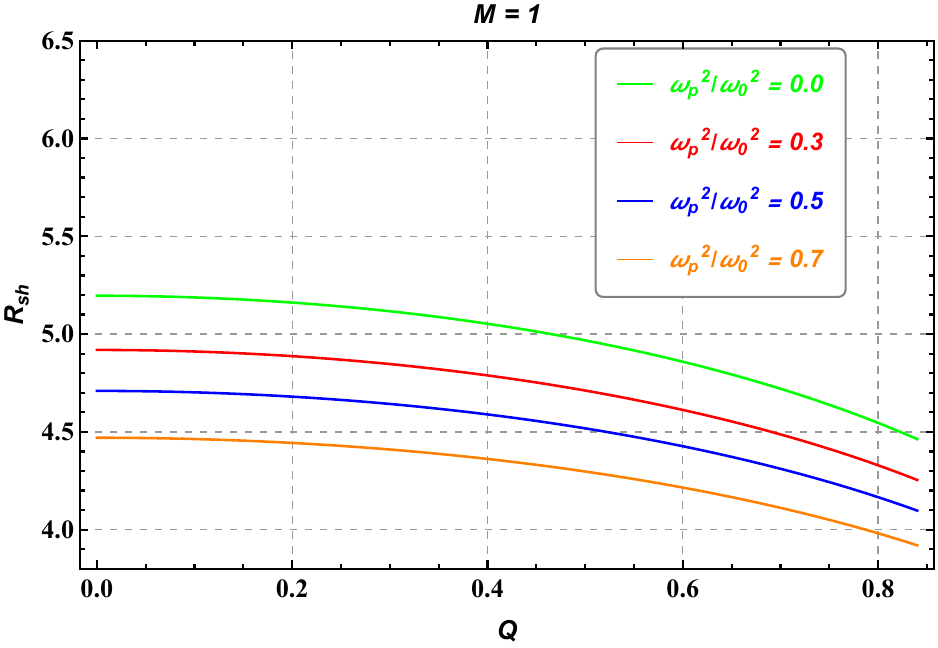}

\caption{Left panel: the plot illustrates the shadow radius as a function of the uniform plasma frequency for the different values of BH charge. The right panel shows the dependence of the shadow radii on the BH charge for the different values of the uniform plasma frequency.} \label{fig.rsh}
\end{figure*}
\begin{figure*}
    \centering
    \includegraphics[scale=0.55]{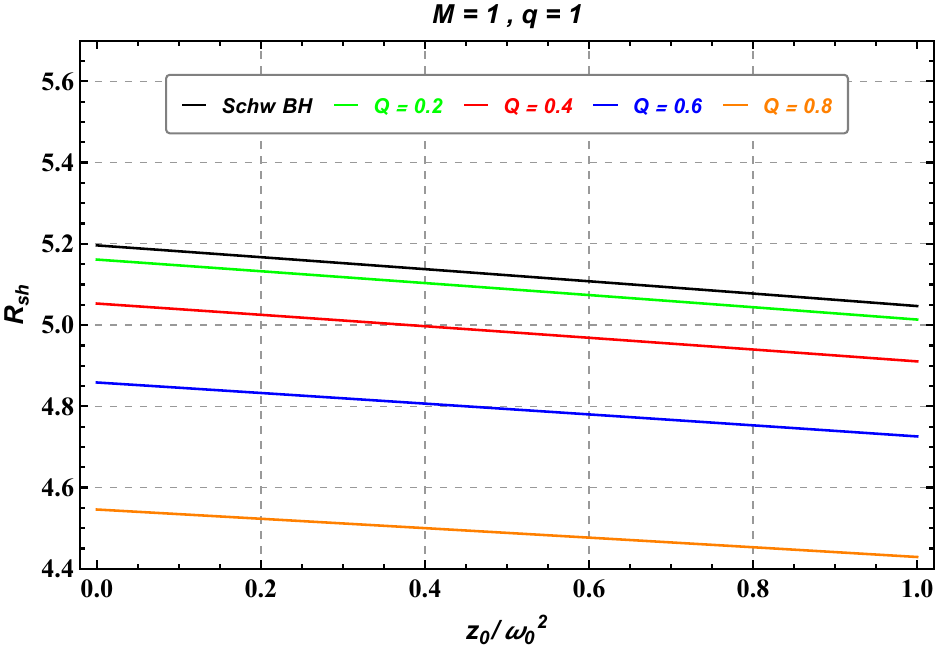}
    \includegraphics[scale=0.55]{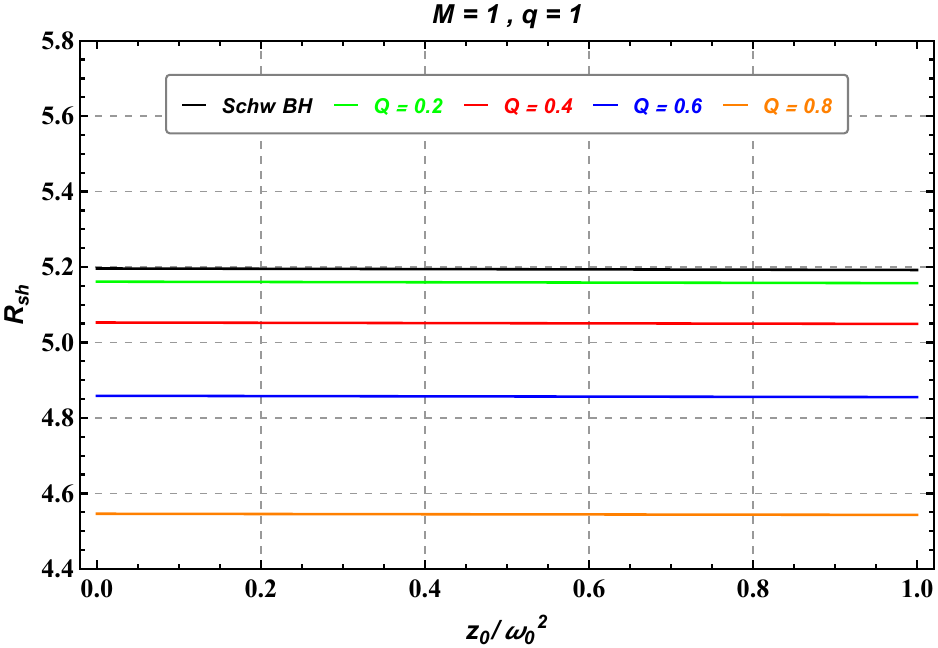}
    \centering
    
    \includegraphics[scale=0.55]{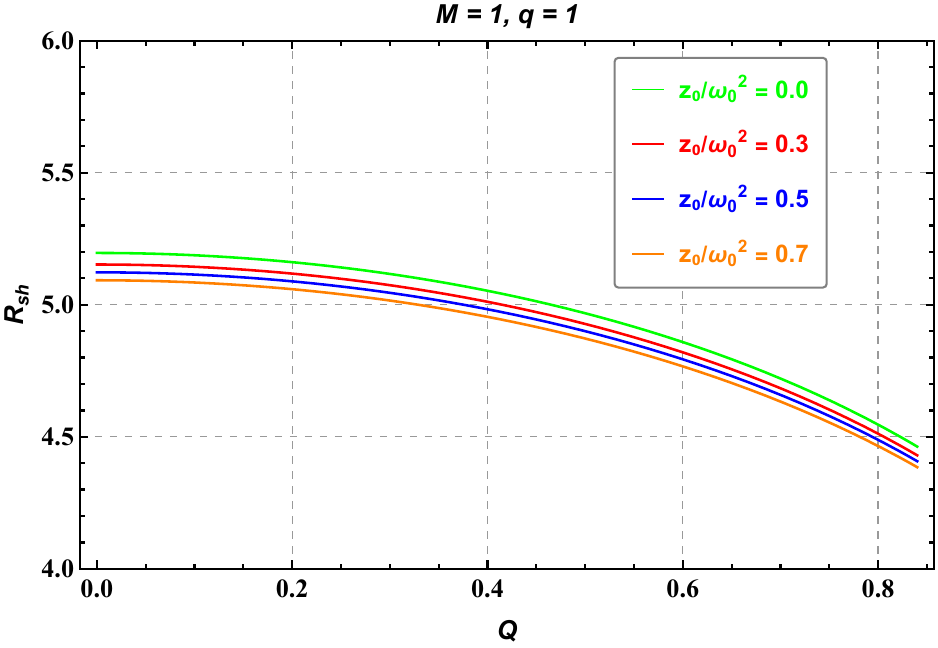}
    \includegraphics[scale=0.55]{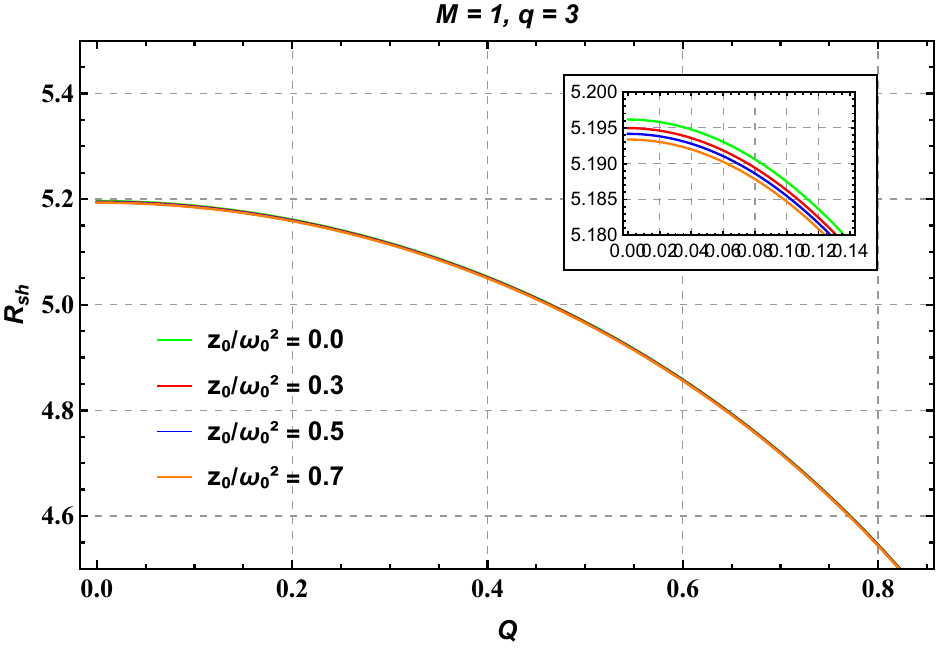}
    \caption{Top panels show the dependence of the shadow radii on the non-uniform plasma frequency for the various values of BH charge, while bottom panels demonstrate shadow radii as a function of BH charge, $Q$, for the different values of non-uniform plasma frequency. The left and right panels, respectively, correspond to $q=1$ and $q=3$.} \label{fig.nonrsh}
\end{figure*}
In the asymptotic region, where $r\to\infty$, the metric function satisfies $A(r)\to 1$. At the same time, $\omega(\infty)=\omega_0=-p_t$ \cite{Perlick_2015}. Thus, $\omega_0$ represents the photon energy or frequency observed at spatial infinity. Using the constraint $\mathcal{H}=0$, we obtain:

 \begin{equation}
     \frac{\omega_0^2}{A(r)}>\omega_p^2(r)
 \end{equation}

This condition means that the local photon frequency, $\omega(r)$, must be higher than the plasma frequency at the same radius. In this case, photons can move through the plasma. Therefore, the plasma can change the apparent size of the BH shadow compared with the vacuum case, where $\omega_p=0$. Using Eq.~\eqref{eqn}, Eq.~\eqref{eqH} can be written as follows 

\begin{equation}\label{Hplasma}
    \mathcal{H}=\frac{1}{2}\left[ g^{\alpha\beta}p_\alpha p_\beta+\omega^2_p \right]\ .
\end{equation}
Photon trajectories are determined from the Hamiltonian equation, 
$\dot{x}^{\alpha}=\partial \mathcal{H}/\partial p_{\alpha}$. 
Considering photon motion in the equatorial plane, where $\theta=\pi/2$ and 
$p_{\theta}=0$, we obtain the following null geodesic equations:

\begin{equation}
    \dot{t}=\frac{dt}{d\lambda}=-\frac{p_{t}}{A(r)}\ ,
\end{equation}
\begin{equation}
    \dot{r}=\frac{dr}{d\lambda}=\frac{p_{r}}{B(r)}\ , \label{eq.33}
\end{equation}
\begin{equation}
    \dot{\phi}=\frac{d\phi}{d\lambda}=\frac{p_{\phi}}{r^2}\ . \label{eq.34}
\end{equation}
Using the ratio of equations Eqs.(\ref{eq.33}) and (\ref{eq.34}), we write the orbital equation as follows:

\begin{equation}
    \frac{dr}{d\phi}=\frac{g^{rr}p_r}{g^{\phi \phi} p_{\phi}}\ .
\end{equation}
For photon geodesics with $\mathcal{H}=0$, the above equation takes the following form:
\begin{equation}
    \frac{dr}{d\phi}=\sqrt{\frac{g^{rr}}{g^{\phi \phi}}} \sqrt{\gamma^2(r)\frac{\omega^2_0}{p^2_{\phi}}-1}\ ,\label{radialPlasma}
\end{equation}
where the function $\gamma^2(r)$ is given by~\cite{Perlick_2015}

\begin{equation}
    \gamma^2(r)=-\frac{g^{tt}}{g^{\phi \phi }}-\frac{\omega^2_{p}}{g^{\phi \phi}\omega^2_0}\ . \label{eq.37}
\end{equation}
A photon arriving from infinity reaches its minimum radial distance at  $r=r_{\rm ph}$ and then moves back to infinity. At this point,  $\gamma^2(r)$ has a turning point. According to this, the photon-sphere radius is  obtained by solving:

\begin{figure*}  
\includegraphics[scale=0.6]{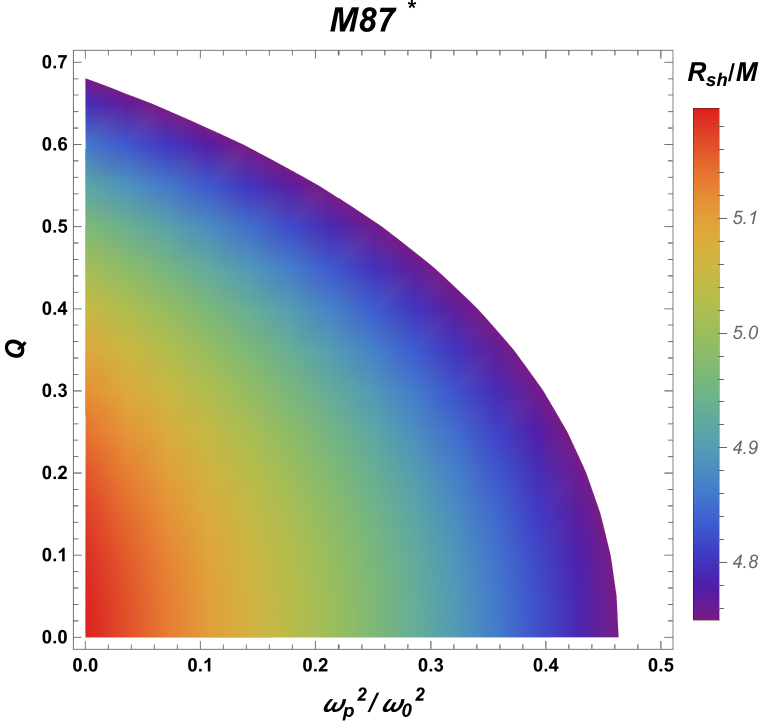}
\includegraphics[scale=0.6]{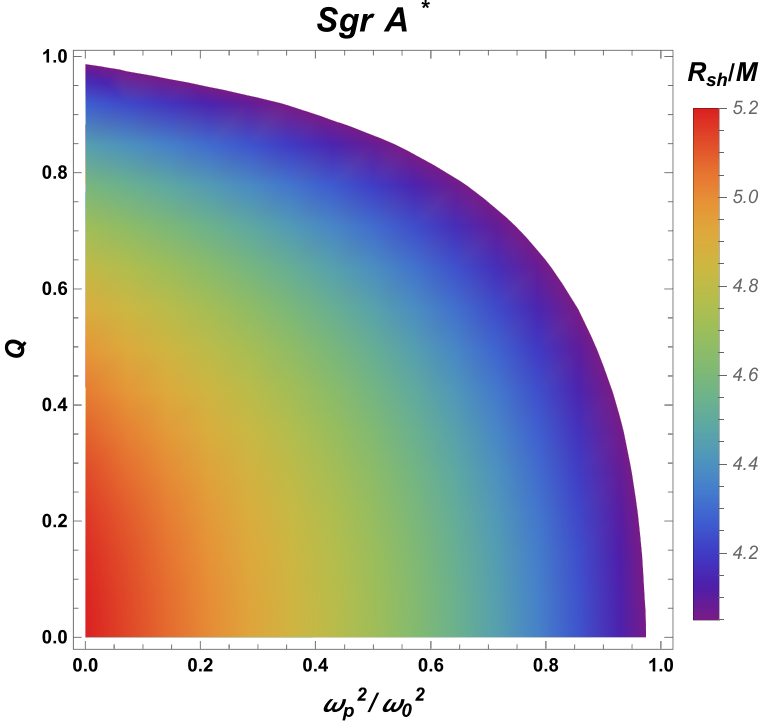}  
\caption {Allowed ranges of the BH charge as a function of the plasma parameter $\omega_{p}^2$/$\omega_{0}^2$ for the SMBHs M87* and Sgr A*.}
\label{fig.SMBH}
\end{figure*}

\begin{equation}
\left. \frac{d\big(\gamma^{2}(r)\big)}{dr} \right|_{r = r_{\text{ph}}} = 0 .
\label{eq.grphoton}
\end{equation}
The photon-sphere radius is obtained numerically for different values of the BH charge $Q$, as shown in Fig.~\ref{fig1}. The figure shows that increasing $Q$ leads to a smaller photon-sphere radius. A similar decrease is also observed for the event horizon and shadow radius. To study the apparent size of the BH shadow, we introduce its angular radius $\alpha_{\rm sh}$, which is given by~\cite{Perlick_2015, KONOPLYA20191}:

\begin{align}
\sin^{2} \alpha_{sh} &= \frac{\gamma^{2}(r_{ph})}{\gamma^{2}(r_{o})}= \notag \\[4pt]
&= \frac{
r_{ph}^{2} \left[ \frac{1}{A(r_{ph})} - \frac{\omega_{p}^{2}(r_{ph})}{\omega_{0}^{2}} \right]
}{
r_{o}^{2} \left[ \frac{1}{A(r_{o})} - \frac{\omega_{p}^{2}(r_{o})}{\omega_{0}^{2}} \right]
}\ ,
\end{align}
In this expression, $r_{o}$ represents the position of the observer, and $r_{\mathrm{ph}}$ represents the position of the photon sphere. When the BH is very far from the observer, the radius of the BH shadow can be approximated using the following equation\cite{KONOPLYA20191}.

\begin{align}
R_{sh} &\simeq r_{o} \sin\alpha_{sh}= \notag\\
&= \sqrt{ r^2_{ph} \left[ \frac{1}{A(r_{ph})} - \frac{\omega^2_{p}(r_{ph})}{\omega^2_0} \right] }\ .
\end{align}

For the plasma models considered here, Eq.~\eqref{eq.37} shows that \(\gamma(r)\) approaches \(r\) as \(r\to\infty\). This behavior allows us to use the above approximation.

Fig.~\ref{fig.rsh}. shows the shadow radius of the BH in a uniform plasma. The left panel shows that the shadow radius becomes smaller as the plasma parameter $\omega_p^2/\omega_0^2$ increases. For a fixed plasma parameter, increasing the BH charge $Q$ also reduces the shadow radius. As shown in the right panel, both the electric charge and the plasma have a decreasing effect on the apparent size of the BH shadow.

In this section, we also studied the BH shadow radius under conditions of a non-uniform plasma medium. We assume that the plasma frequency follows a simple exponential profile of the form\cite{Rogers2015,2018MNRAS.475..867E}:
\begin{eqnarray}
    \omega_p^2(r)=\frac{z_0}{r^q}
\end{eqnarray}
where $z_0$ and $q>0$ denote the free parameters of the plasma distribution. To examine the main features of this model, we consider two specific cases:

\begin{itemize}
\item $q=1$ with constant $z_0$: This case reproduces the negative-mass diverging lens model~\cite{2018MNRAS.475..867E}.

\item $q=3$ with constant $z_0$: This case describes a plasma profile related to the Goldreich--Julian (GJ) density near the stellar surface.
\end{itemize}

Fig.~\ref{fig.nonrsh}. shows the shadow radius of the BH in a non-uniform plasma. For $q=1$, increasing the plasma parameter leads to a small decrease in the shadow radius, while increasing the charge $Q$ produces a stronger reduction. For $q=3$, the curves corresponding to different plasma parameters are almost the same, showing that the plasma effect is very weak. In this case, the change in the shadow radius is mainly caused by the BH charge.

In our analysis, $M87^*$ and $Sgr~A^*$ are modeled as static and spherically symmetric BHs. We use the EHT measurements of the angular diameter, BH mass, and distance from Earth to constrain the charge $Q$ and the plasma parameter $\omega_p^2/\omega_0^2$. 

\begin{table}[htbp]
\centering
\caption{Observational Data for M87* and Sgr A*~\cite{EHT2024,GRAVITY:2020gka,delaurentis2022accuratemassdistributionm87}.}

\resizebox{0.5\textwidth}{!}{
\begin{tabular}{|l|c|c|}
\hline
\textbf{Parameter}         & \textbf{M87*} & \textbf{Sgr A*} \\ \hline
Angular Diameter($\theta$) & $43.3 \pm 2.3 \, \mu\text{as}$ & $51.8 \pm 2.3 \, \mu\text{as}$ \\ \hline
Distance ($D$)             & $16.5  \, \text{Mpc}$ & $8.275   \, \text{kpc}$ \\ \hline
Mass ($M$)                 & $(6.5 \pm 0.7) \times 10^9 \, M_{\odot}$ & $(4.297 \pm 0.013) \times 10^6 \, M_{\odot}$ \\ \hline
\end{tabular}
} 

\label{tab:data}
\end{table}

These observational quantities are listed in Table~\ref{tab:data}. The corresponding allowed regions are presented in Fig.~\ref{fig.SMBH}. For both sources, a larger plasma parameter allows a smaller range of the BH charge. Using the observational data, the shadow diameter normalized by the BH mass can be calculated from the relation~\cite{Bambi_2019}

\begin{equation}
    d_{sh}=\frac{D\theta}{M}\, .
\end{equation}
We can write the shadow diameter as follows based on the observational data presented in Table~\ref{tab:data}

\begin{equation}
\begin{aligned}
d_{\mathrm{sh}}^{M87^*} &= (11.1 \pm 1.3)\,M, \\
d_{\mathrm{sh}}^{SgrA^*} &= (10.1 \pm 0.4)\,M.
\end{aligned}
\end{equation}
The relation $d_{\rm sh}=2R_{\rm sh}$ allows us to obtain constraints on the BH charge $Q$ and the plasma parameter $\omega_p^2/\omega_0^2$. The corresponding allowed parameter regions are presented in Fig.~\ref{fig.SMBH}. The left panel shows the results for $M87^*$, while the right panel corresponds to $Sgr~A^*$. 

\section{Gravitational lensing under weak-field conditions in a plasma environment}
\label{sec:weak}

In this section we investigate the gravitational lensing in the spacetime of a charged quantum-corrected BH. To obtain the weak gravitational lensing effects, the metric is considered in the weak-field approximation. (see, e.g.,~\cite{Bisnovatyi-Kogan2010,2021PDU....3200798B})

\begin{equation}
    g_{\alpha \beta}=\eta_{\alpha \beta}+h_{\alpha \beta}\, .
\end{equation}
In this expression, $\eta_{\alpha\beta}$ denotes the gravitational potential associated with the Minkowski spacetime metric, while $h_{\alpha\beta}$ describes the perturbation due to the gravitational field. Based on the above considerations and following~\cite{Bisnovatyi-Kogan2010}, the spacetime metric can be written as follows:

\begin{eqnarray}
 &&   \eta_{\alpha \beta}=diag(-1,1,1,1)\ , \nonumber\\
 &&   h_{\alpha \beta} \ll 1, \hspace{0.5cm} h_{\alpha \beta} \rightarrow 0 \hspace{0.5cm} under\hspace{0.2cm}  x^{\alpha}\rightarrow \infty \ ,\nonumber\\
 &&     g^{\alpha \beta}=\eta^{\alpha \beta}-h^{\alpha \beta}, \hspace{0,5cm} h^{\alpha \beta}=h_{\alpha \beta}\, .
\end{eqnarray}
Using this relation, we can calculate the light deflection angle in the spacetime of a charged quantum-corrected BH as follows:

\begin{eqnarray}
    \hat{\alpha }_{\text{b}}&=&\frac{1}{2}\int_{-\infty}^{\infty}\frac{b}{r}\left(\frac{dh_{33}}{dr}+\frac{1}{1-\omega^2_p/ \omega_0^2}\frac{dh_{00}}{dr}-\right.\nonumber\\&-&\left.\frac{K_e}{\omega_0^2-\omega^2_p}\frac{dN}{dr} \right)dz\, , 
\end{eqnarray}
The quantities $\omega_{p}$ and $\omega_0$ represent the plasma and photon frequencies, respectively. Using the weak-field approximation, the spacetime metric can be expanded in a Taylor series as follows:

\begin{equation}
\begin{aligned}
ds^2={}&ds_0^2
+\left(
\frac{2M}{r}-\frac{Q^2}{r^2}
\right)dt^2+
\\
&+\frac{1}{r^2}
\left[
\frac{M(2\Delta+1)}{\Delta}\,r
+\frac{1}{2}+3M^2+\Delta+
\right.
\\
&\left.
\qquad
+\frac{M(2\sqrt{2}M-1)}{2\Delta}
+\frac{M^2}{\Delta^2}
\right]dr^2 .
\end{aligned}
\label{eq:metric_33}
\end{equation}
where $ds^2_0=-dt^2+dr^2+r^2(d\theta^2+\sin^2\theta d\phi^2)$ and, to simplify the expression, we introduced the following notation: $\Delta = \sqrt{2}M-Q^2$. The corresponding components of the metric perturbation $h_{\alpha \beta}$ in Cartesian coordinates are given by:

\begin{eqnarray}
    h_{00}=\frac{2M}{r}-\frac{Q^2}{r^2}\,,
\end{eqnarray}
\begin{equation}
\begin{aligned}
h_{ik}={}&
\frac{1}{r^2}
\left[
\frac{M(2\Delta+1)}{\Delta}\,r
+\frac{1}{2}+3M^2+\Delta+
\right.
\\
&\left.
\qquad
+\frac{M(2\sqrt{2}M-1)}{2\Delta}
+\frac{M^2}{\Delta^2}
\right] n_i n_k .
\end{aligned}
\label{hik}
\end{equation}
\begin{equation} \label{h33}
\begin{split}
h_{33} ={}&
\frac{1}{r^2}
\left[
\frac{M(2\Delta+1)}{\Delta}\,r
+\frac{1}{2}+3M^2+\Delta+
\right.
\\
&\left.
\qquad
+\frac{M(2\sqrt{2}M-1)}{2\Delta}
+\frac{M^2}{\Delta^2}
\right]\cos^2\chi \,.
\end{split}
\end{equation}
where  $\Delta = \sqrt{2}M-Q^2$ , $\cos^2\chi=z^2/(b^2+z^2)$ and $r^2=b^2+z^2$. The total deflection angle can then be written as the sum of three contributions, as follows:
\begin{equation} \label{ab}
\hat{\alpha}_{b}=\hat{\alpha}_{1}+\hat{\alpha}_{2}+\hat{\alpha}_{3}\ ,
\end{equation}
with

\begin{equation} 
\begin{aligned}\label{a1a2a3}
   \hat{\alpha}_{1} &= \frac{1}{2}\int_{-\infty}^{\infty} \frac{b}{r}\frac{dh_{33}}{dr} \, dz,\\ 
   \hat{\alpha}_{2} &= \frac{1}{2}\int_{-\infty}^{\infty} \frac{b}{r}\frac{1}{1-\omega^2_{p}/\omega_0^2}\frac{dh_{00}}{dr} \, dz,\\ 
   \hat{\alpha}_{3} &= \frac{1}{2}\int_{-\infty}^{\infty} \frac{b}{r}\Bigg(-\frac{K_{e}}{\omega_0^2-\omega^2_{p}}\frac{dN}{dr}\Bigg) \, dz.
\end{aligned}
\end{equation}

In the following subsections, we study the light deflection angle for two plasma models: uniform and non-uniform plasma distributions.

\subsection{Uniform Plasma}

In the case of  uniform plasma, Eq.~\eqref{ab} takes the following form:~\cite{Khasanov2025JCAP,Alloqulov12025,Alloqulov2024}:
\begin{equation}\label{aunihat} \hat{\alpha}_{uni}=\hat{\alpha}_{uni1}+\hat{\alpha}_{uni2}+\hat{\alpha}_{uni3}\ ,
\end{equation}

\begin{figure*}[!htb]
\centering
\includegraphics[scale=0.5]{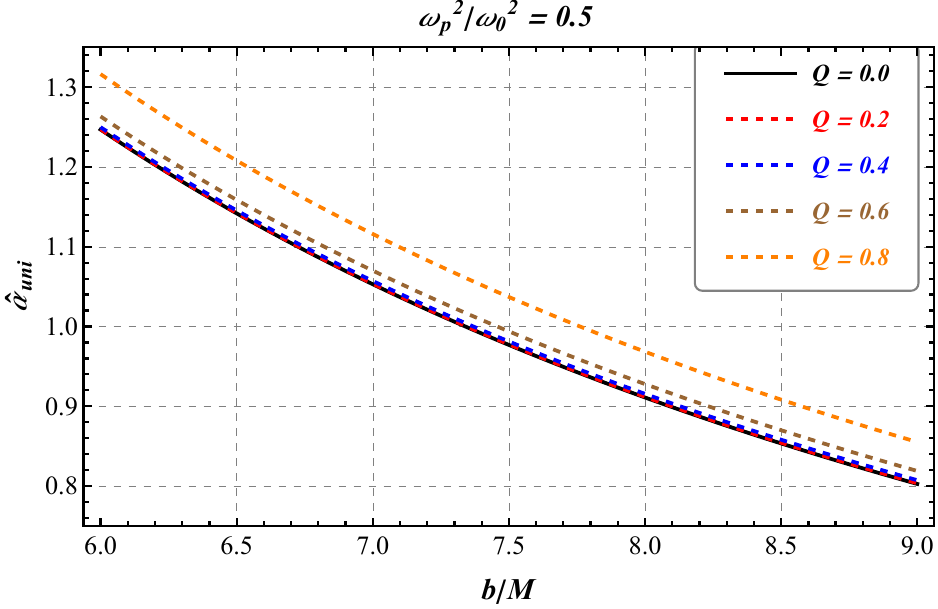}
\includegraphics[scale=0.5]{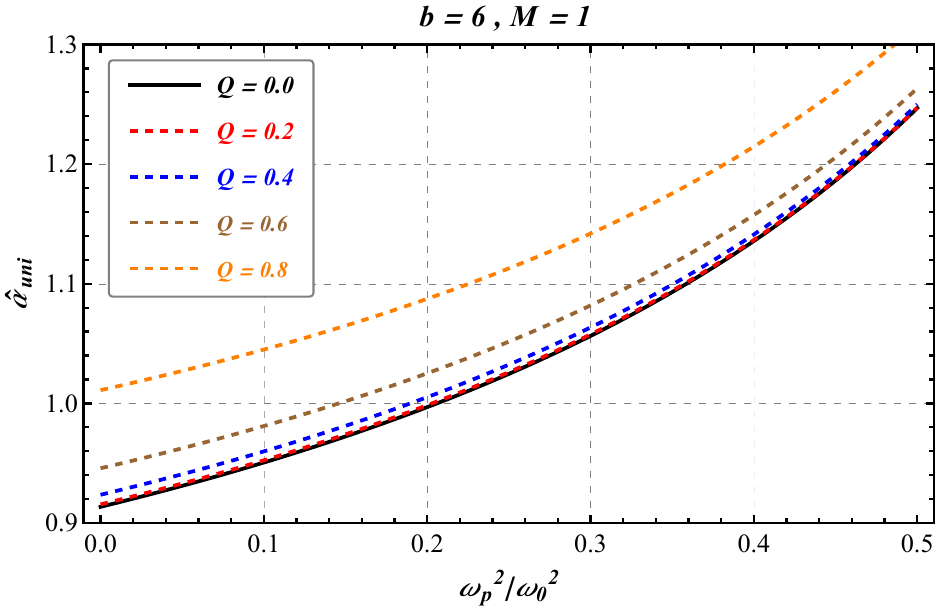}
\caption{The left panel represents the deflection angle $\hat{\alpha}_{\rm uni}$ as a function of the impact parameter for the different values of the BH charge. Here, we set $\omega^2_p/\omega_0^2=0.5$. Right panel: the dependence of the deflection angle $\hat{\alpha}_{\rm uni}$ on $\omega^2_p/\omega_0^2$ for the different values of the BH charge. $b/M=6$ was considered for this panel. }\label{fig:auni}
\end{figure*}

Using Eqs.~\eqref{a1a2a3} and ~\eqref{aunihat}, the deflection angle of light is obtained as:

\begin{equation}
\begin{aligned}
\hat{\alpha}_{\rm uni}
={}&
\frac{M(2\Delta+1)}{\Delta b}
+\frac{\pi}{8\Delta^2 b^2}
\Big[
2\Delta^3
+(1-2\sqrt{2}M)\Delta^2+
\\
&\qquad
+\sqrt{2}M(16M^2-1)\Delta
+2M^2(1+2\sqrt{2}M-8M^2)
\Big]+
\\
&+
\left(
\frac{2M}{b}
-\frac{\pi Q^2}{2b^2}
\right)
\frac{\omega_0^2}{\omega_0^2-\omega_e^2}.
\end{aligned}
\end{equation}
Based on the obtained expression, we study the light deflection angle $\hat{\alpha}_{\rm uni}$ in a uniform plasma for different values of the  BH charge $Q$, as shown in Fig.~\ref{fig:auni}. In the left panel, $\hat{\alpha}_{\rm uni}$ decreases with increasing impact parameter $b/M$,  while a larger charge leads to a stronger deflection. In the right panel, 
the deflection angle increases as the plasma parameter 
$\omega_p^2/\omega_0^2$ increases. The effect of the BH charge is also  clear, since larger values of $Q$ shift the deflection angle to higher values.

\subsection{Non uniform plasma}

In this subsection of our analysis, we use the singular isothermal sphere (SIS) model to study the weak deflection of photons in BH spacetime. Typically, a SIS is a spherical gas cloud with a central singularity whose density approaches infinity. The density distribution of the SIS can be written as: \cite{Bisnovatyi-Kogan2010,2021PDU....3200798B}
\begin{equation}
\rho(r)=\frac{\sigma^2_{\nu}}{2\pi r^2}\ ,
\end{equation}
The quantity $\sigma_\nu^2$ represents the one-dimensional velocity dispersion, and the plasma concentration can be written in the following analytical form:
\begin{equation}
    N(r)=\frac{\rho(r)}{km_p}  \,.
\end{equation}
Commonly, the above expressions are associated with the plasma frequency as follows:

 \begin{equation}
\omega^2_e=K_e N(r)=\frac{K_e \sigma^2_{\nu}}{2\pi \kappa m_p r^2}\ .
\end{equation}
After that, we investigate the effect of non-uniform plasma (SIS) on the deflection angle around the spacetime of a charged quantum-corrected BH using the following expression:
\begin{equation} \label{asis}\hat{\alpha}_{SIS}=\hat{\alpha}_{SIS1}+\hat{\alpha}_{SIS2}+\hat{\alpha}_{SIS3}\ ,
\end{equation}
Using equations (\ref{h33}), (\ref{a1a2a3}) and (\ref{asis}), the deflection angle can be written in the following form:
\begin{equation}
\begin{aligned}
\hat{\alpha}_{\rm SIS}
={}&
\frac{M(2\Delta+1)}{\Delta b}
+\frac{\pi}{8\Delta^2 b^2}
\Big[
2\Delta^3+(1-2\sqrt{2}M)\Delta^2+
\\[-1.5mm]
&\hspace{-1.2em}
+\sqrt{2}M(16M^2-1)\Delta
+2M^2(1+2\sqrt{2}M-8M^2)
\Big]-
\\[-1.5mm]
&\hspace{-1.2em}
-\left(
\frac{\pi Q^2}{2b^2}
-\frac{2M}{b}
-\frac{16M^3\omega_c^2}{3\pi b^3\omega^2}
+\frac{3M^2Q^2\omega_c^2}{2b^4\omega^2}
\right)
-\frac{4M^2\omega_c^2}{\pi b\omega^2}.
\end{aligned}
\end{equation}

These calculations introduce an additional plasma parameter $\omega_c^2$, which can be written as follows:
 \begin{equation}
\omega^2_c=K_e N(r)=\frac{K_e \sigma^2_{\nu}}{2\pi \kappa m_p R_S^2}\ .
\end{equation}

\begin{figure*}
    \centering
    \includegraphics[scale=0.5]{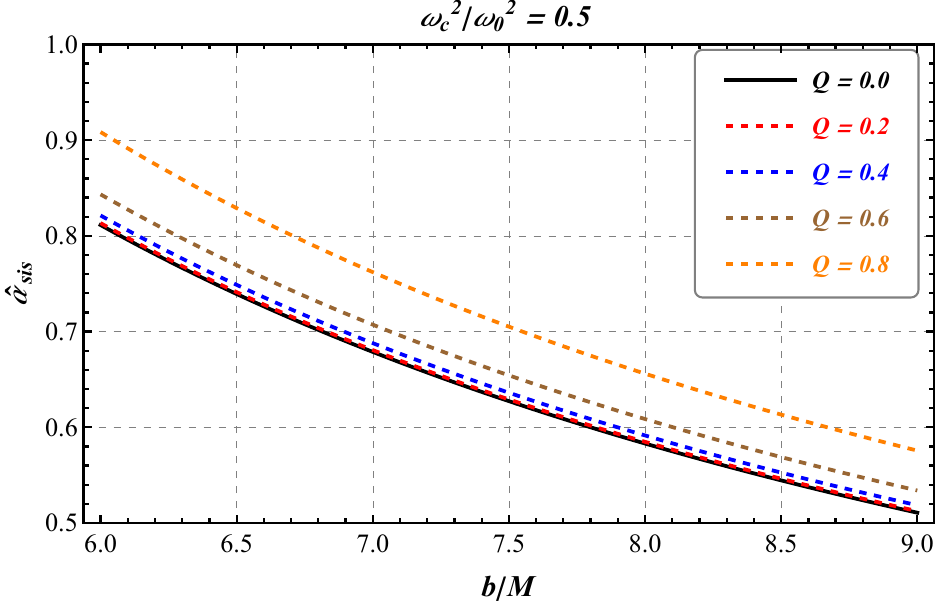}
    \includegraphics[scale=0.5]{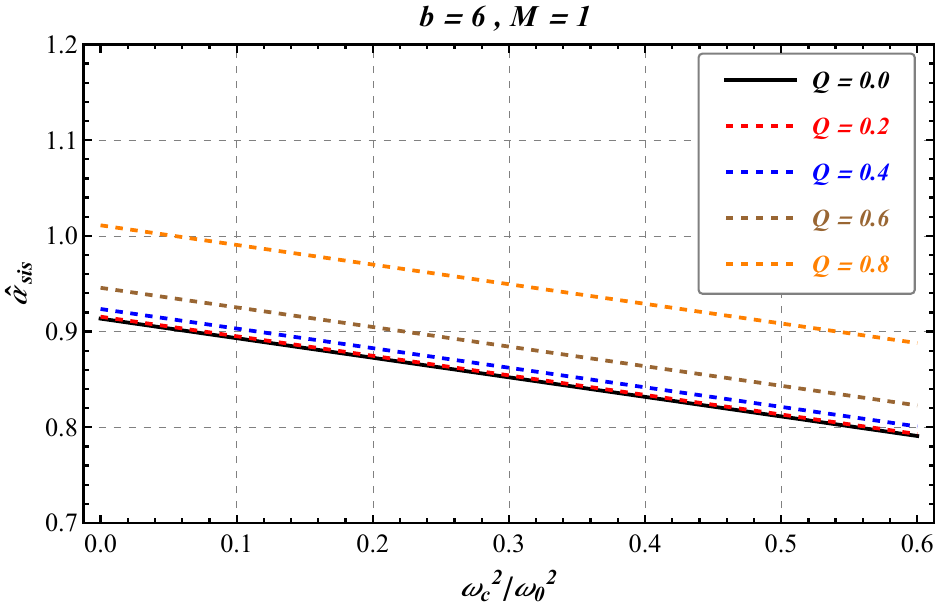}
    \caption{The deflection angle $\hat{\alpha}_{\rm SIS}$ for the non-uniform plasma distribution. The left panel shows its dependence on the impact parameter $b/M$ for different values of the BH charge, while the right panel illustrates its variation with the normalized plasma parameter $\omega_c^2/\omega_0^2$ at $b/M=6$.} \label{Asis}
\end{figure*}

Based on the obtained expression, we study the light deflection angle 
$\hat{\alpha}_{\rm SIS}$ in a non-uniform plasma for different values of the BH charge $Q$, as shown in Fig.~\ref{Asis}. In the left panel, the deflection angle decreases as the impact parameter $b/M$ increases, while larger values of $Q$ lead to a stronger deflection of light. The right panel shows that $\hat{\alpha}_{\rm SIS}$ decreases as the plasma parameter increases. At the same time, increasing the BH charge shifts the deflection angle to larger values. This behavior is similar to that found for the uniform plasma case. However, the magnitude of the deflection angle in the non-uniform plasma is smaller than that obtained for the uniform plasma.

\begin{figure*}[!htb]
\centering    
\includegraphics[scale=0.4]{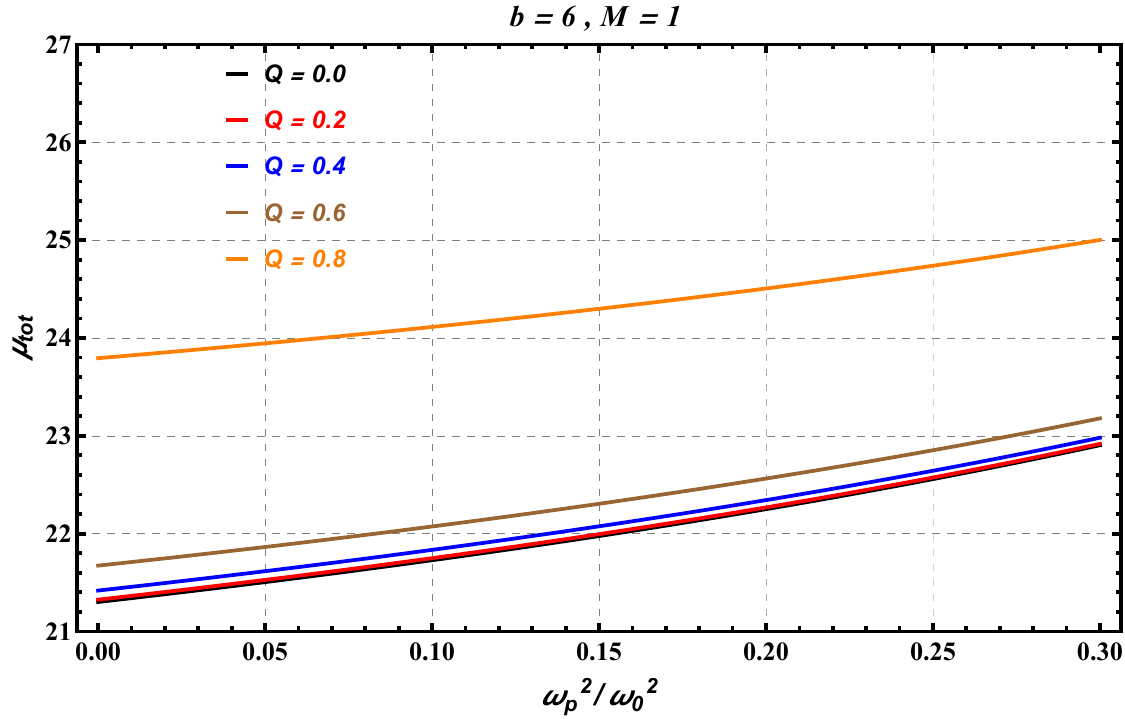}
\includegraphics[scale=0.48]{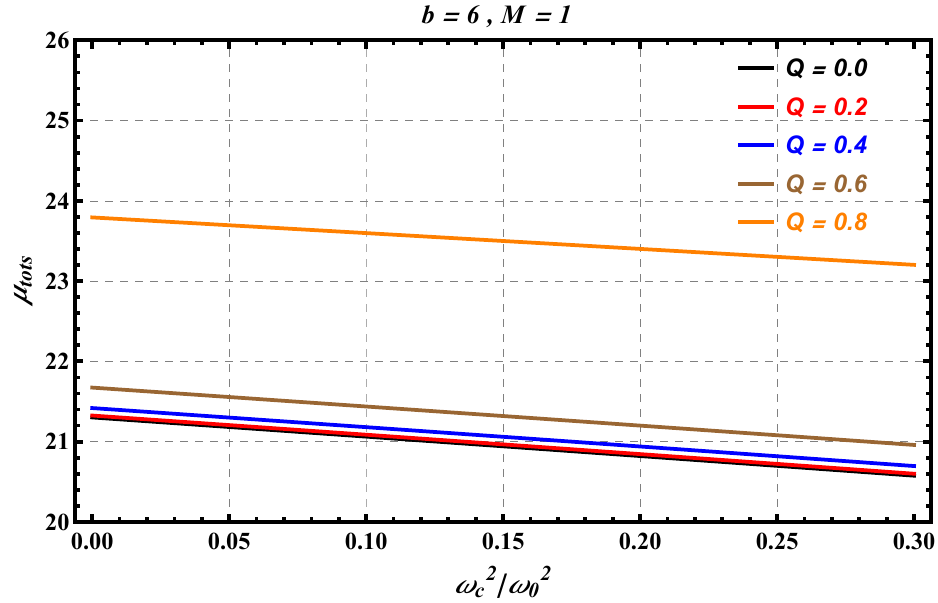}      
\caption{Dependence of the total magnification on the normalized plasma frequency for different values of the BH charge $Q$. The left panel corresponds to uniform plasma, while the right panel shows the result for non-uniform plasma.}
\label{fig:mu}
\end{figure*}

\section{MAGNIFICATION OF GRAVITATIONALLY LENSED  IMAGES}\label{sec:mag}

We now investigate the brightness of images produced by gravitational lensing in a plasma around a charged quantum-corrected black hole. By using the lens equation, the image brightness can be related to the deflection angle $\hat{\alpha}$, the image angle $\theta$, and the source angle $\beta$~\cite{Bozza2008lens}.
\begin{align}\label{lenseq}
\theta D_\mathrm{s}=\beta D_\mathrm{s}+\hat{\alpha}D_\mathrm{ds}\, . 
\end{align}
Where $D_\mathrm{s}$, $D_\mathrm{d}$, and $D_\mathrm{ds}$ represent the distances between the source and observer, the lens and observer, and the source and lens, respectively, while $\theta$ and $\beta$ represent the angular positions of the image and the source. Based on the above equation, we can write the following expression for $\beta$:
\begin{align}\label{newlenseq}
\beta=\theta -\frac{D_\mathrm{ds}}{D_\mathrm{s}}\frac{\xi(\theta)}{D_\mathrm{d}}\frac{1}{\theta}\,.
\end{align}
In this relation, $\xi(\theta)=|\hat{\alpha}_b|b$ and $b=D_{\mathrm d}\theta$. If the lensed image forms a circular ring,  it is called an Einstein ring. Its radius is given by  $R_s=D_{\mathrm d}\theta_E$, where $\theta_E$ denotes the angular  Einstein radius. For the vacuum case, the value of $\theta_E$ follows  from the spacetime geometry and the lensing distances, and is given by~\cite{1999grle.book.....S}:
 \begin{figure*}
         \centering
         \includegraphics[scale=0.35]{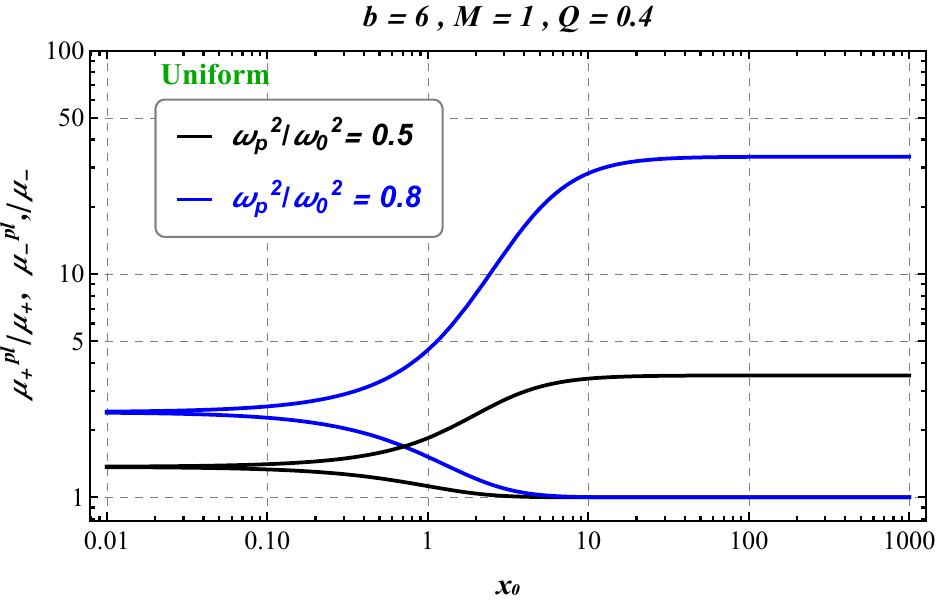}
         \includegraphics[scale=0.35]{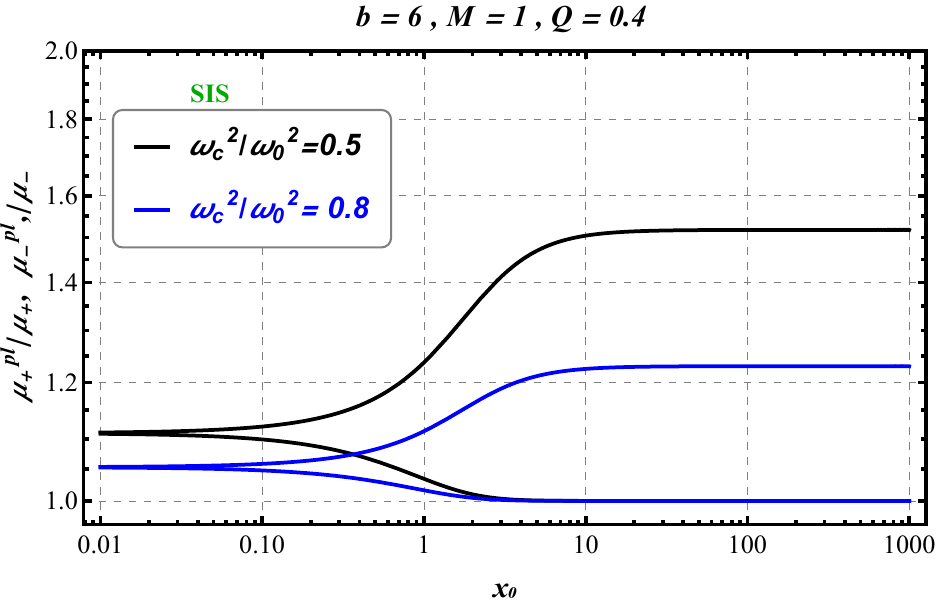}
         \includegraphics[scale=0.35]{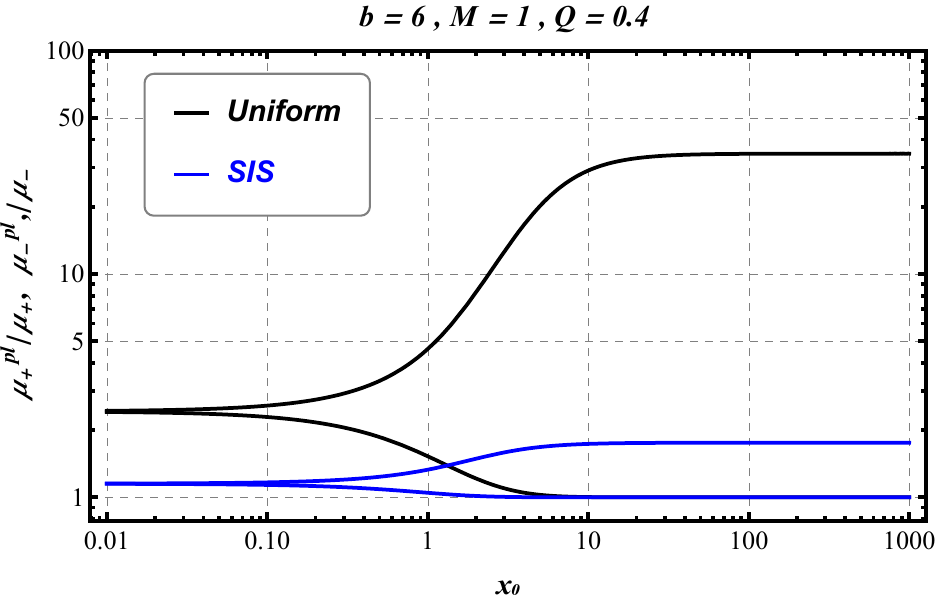}
         \caption{The plot demonstrates the magnification ratio of uniform and non-uniform plasma to vacuum for different values of
the plasma frequency.
}\label{myux}
     \end{figure*}

     \begin{eqnarray}
\theta_E=\sqrt{2R_s\frac{D_{ds}}{D_dD_s}}\ .
\end{eqnarray}
Now we study the equation for the magnification of brightness.
\begin{align}\label{magni}
\mu_{\Sigma}=\frac{I_\mathrm{tot}}{I_*}=\underset{k}\sum\bigg|\bigg(\frac{\theta_k}{\beta}\bigg)\bigg(\frac{d\theta_k}{d\beta}\bigg)\bigg|, \quad k=1,2, \cdot \cdot \cdot ,  j\ ,
\end{align}
where $I_*$ and $I_{\mathrm{tot}}$ represent the unlined brightness of the source and the combined brightness of all images, respectively. The magnification of the source can be expressed as follows.
\begin{align}\label{postparity}
 \mu^\mathrm{pl}_\mathrm{+}=\frac{1}{4}\bigg(\frac{x}{\sqrt{x^2+4}}+\frac{\sqrt{x^2+4}}{x}+2\bigg) ,
\end{align}

\begin{align}\label{negparity}
\mu^\mathrm{pl}_\mathrm{-}=\frac{1}{4}\bigg(\frac{x}{\sqrt{x^2+4}}+\frac{\sqrt{x^2+4}}{x}-2\bigg).
\end{align}
In this expression, $x={\beta}/{\theta_0}$ is a dimensionless parameter that represents individual images $\mu_{+}^{\mathrm{pl}}$ and $\mu_{-}^{\mathrm{pl}}$. Using equations (\ref{postparity}) and (\ref{negparity}), we can express the full magnification as follows:
\begin{align}\label{magtot}
\mu^\mathrm{pl}_\mathrm{tot}=\mu^\mathrm{pl}_{+}+\mu^\mathrm{pl}_{-}=\frac{x^2+2}{x\sqrt{x^2+4}}\ .
\end{align}
We now investigate the image magnification produced by gravitational lensing around a charged quantum-corrected BH surrounded by plasma. We consider both uniform and non-uniform plasma distributions.

\textit{Uniform Plasma case}: we examine how a uniform plasma affects magnification in a charged quantum-corrected black hole environment. For this purpose, we first define the total magnification and angle (i.e., $\mu^{pl}_{tot}$ and $\theta^{pl}_{uni}$) as follows:
\begin{equation}
    \mu^{pl}_{tot}=\mu^{pl}_{+}+\mu^{pl}_{-}=\dfrac{x^2_{uni}+2}{x_{uni}\sqrt{x^2_{uni}+4}} ,
\end{equation}
Here, the quantities of $(\mu^{pl}_+)_{uni}$ and $(\mu^{pl}_-)_{uni}$ can be written as follows:
\begin{eqnarray}
       (\mu^{pl}_+)_{uni}&=&\frac{1}{4}\left(\dfrac{x_{uni}}{\sqrt{x^2_{uni}+4}}+\dfrac{\sqrt{x^2_{uni}+4}}{x_{uni}}+2\right) ,\\
       (\mu^{pl}_-)_{uni}&=&\frac{1}{4}\left(\dfrac{x_{uni}}{\sqrt{x^2_{uni}+4}}+\dfrac{\sqrt{x^2_{uni}+4}}{x_{uni}}-2\right) .
   \end{eqnarray}
\textit{Non-uniform Plasma case}: We now examine how a non-uniform plasma (singular isothermal sphere (as SIS medium)) affects the magnification of lensed images,  which can be defined by
\begin{equation}
    (\mu^{pl}_{tot})_{SIS}=(\mu^{pl}_{+})_{SIS}+(\mu^{pl}_{-})_{SIS}=\dfrac{x^2_{SIS}+2}{x_{SIS}\sqrt{x^2_{SIS}+4}}\ ,\label{eq:magsis}
\end{equation}
with
 \begin{equation}
       (\mu^{pl}_+)_{SIS}=\frac{1}{4}\left(\dfrac{x_{SIS}}{\sqrt{x^2_{SIS}+4}}+\dfrac{\sqrt{x^2_{SIS}+4}}{x_{SIS}}+2\right),\ 
\end{equation}
 \begin{equation}
       (\mu^{pl}_-)_{SIS}=\frac{1}{4}\left(\dfrac{x_{SIS}}{\sqrt{x^2_{SIS}+4}}+\dfrac{\sqrt{x^2_{SIS}+4}}{x_{SIS}}-2\right),\ 
  \end{equation}
 and 
\begin{eqnarray}
   x_{SIS}&=&\frac{\beta}{(\theta^{pl}_E)_{SIS}}\, .  
\end{eqnarray}

Taking the above equations into account, we examine the magnification of the lensed images in both uniform and non-uniform plasma environments. Fig.~\ref{myux} shows the magnification ratio as a function of the normalized source position $x_0$. In the uniform plasma case, the magnification changes more strongly, especially for larger values of the plasma parameter $\omega_p^2/\omega_0^2$. In the SIS plasma case, the magnification changes less with the normalized source position $x_0$. The comparison in the right panel shows that the uniform plasma has a stronger effect on the image magnification. The behavior of the total magnification is further shown in Fig.~\ref{fig:mu}. For the uniform plasma, $\mu_{\rm tot}$ increases as $\omega_p^2/\omega_0^2$ increases, whereas for the SIS plasma, $\mu_{\rm tot}$ decreases with increasing $\omega_c^2/\omega_0^2$. In both plasma models, larger values of the BH charge $Q$ shift the total magnification to higher values.

\section{Conclusions}\label{sec:summary}

In the present work, the optical properties of the BH in EQG have been analyzed. Here, it must be emphasized that incorporation of EQG modifies the upper limit on the BH charge. This indicates that quantum-gravitational effects can influence the admissible parameter space of the charged BH. Consequently, this gives a potential theoretical signature of EQG and has an impact on the observational properties.  Based on our analysis, we summarize our key findings as follows:

First, we concisely reviewed the spacetime of the BH in EQG, including the event horizon structure, and found that the event horizon of this BH shrinks under the effect of the BH charge. Subsequently, the motion of photons around the BH was studied using the Lagrangian formalism. In particular, we obtained the equations of motion and the effective potential for the massless particles. We then explored the radius of the photon sphere by setting the corresponding condition on the effective potential. Our results showed that the radius of the photon sphere decreases with the increase of the BH charge. 

Furthermore, we assumed that the BH in EQG is surrounded by a plasma environment, and we utilized the Hamilton formalism so as to investigate the impact of the plasma on the BH shadow. Two types of plasma: uniform and non-uniform, were considered. We chose the numerical approach to explore it and found that the shadow radius decreases under the influence of the BH charge and uniform plasma frequency. In addition, it was shown that the non-uniform plasma causes the shadow area to shrink slightly. Next, we constrained the BH charge and uniform plasma frequency by applying the EHT observational data for the SMBHs M87* and Sgr A*. 

Further, keeping in mind the assumption that the BH in EQG is surrounded by a uniform and non-uniform plasma, the gravitational weak lensing was explored. In particular, we obtained the deflection angle analytically for the uniform and non-uniform cases. We then demonstrated them as functions of the impact parameter and plasma frequencies. It was found that in the uniform plasma case, the values of the deflection angle increase slightly with the rise of the BH charge and the uniform plasma frequency. In the non-uniform plasma case, there is a rise with the influence of the BH charge, while the angle of deflection decreases with the increase of the non-uniform plasma frequency.  

Finally, the total magnification was evaluated for the same set of plasma environments. We determined that larger values of the BH charge lead to an increase in the total magnification in both cases. Also, the uniform plasma frequency exhibits a positive dependence on the total magnification and vice versa for the non-uniform plasma frequency. The magnification ratios comparing the plasma scenarios with the vacuum case were also presented for the different values of the plasma frequency.   

\section*{ACKNOWLEDGEMENT}

This research was funded by the National Natural Science Foundation of China (NSFC) under Grant No. U2541210.

\bibliography{ref}
\end{document}